\documentclass[12pt]{article}   
\usepackage{epsfig}

\newcommand{\locsection}[1]{\setcounter{equation}{0}\section{#1}}

\def\beq{\begin{equation}}
\def\eeq{\end{equation}}

\def\al{\alpha}

\def\ga{\gamma}

\def\de{\delta}
\def\De{\Delta}

\def\Si{\Sigma}

\def\tm{\times}
\def\La{\Lambda}

\def\sq{\sqrt}

\def\l{\left (}
\def\r{\right )}
\def\fr{\frac}
\def\la{\label}
\def\hs{\hspace}
\def\vs{\vspace}  

\def\ran{\rangle}
\def\lan{\langle}
\def\ov{\overline}
\def\tl{\tilde}

\begin{document}
%\begin{titlepage}

\begin{center}   
{\Large\bf 
Precursors From $S^{(1)}/Z_2\tm Z_2'$ Orbifold GUTs }  
\end{center} 
\vspace{0.5cm} 
\begin{center} 
{\large Filipe \hs{-0.07cm}Paccetti \hs{-0.07cm}Correia$^{a}$, 
{}Michael \hs{-0.07cm}G. \hs{-0.07cm}Schmidt$^{a}$, 
{}Zurab \hs{-0.07cm}Tavartkiladze$^{a, b}$
} 
\vspace{0.5cm}

$^a${\em Institut f\"ur Theoretische Physik, Universit\"at Heidelberg,
Philosophenweg 16,\\
 D-69120 Heidelberg, Germany \\

$^b$ Institute of Physics, 
Georgian Academy of Sciences, Tbilisi 380077, Georgia\\

} 

\end{center}

\vspace{0.5cm}

\begin{abstract}

The possibility of appearance of GUT precursors near the TeV scale 
(suggested by Dienes-Dudas-Gherghetta) is
addressed within  5D GUTs compactified on an $S^{(1)}/Z_2\tm Z_2'$
orbifold. For a low compactification scale (large radius), 
there is a significant non universal logarithmic contribution
in the relative running of gauge couplings. This within 5D $SU(5)$,
with the minimal field content, gives wrong prediction for $\al_3(M_Z)$
unless one goes beyond the minimal setting.
The realization of the light precursors' idea thus requires some specific
extensions. As a scenario alternative to $SU(5)$ we also consider an
$SU(6)$ orbifold GUT, whose minimal non SUSY version gives natural 
unification. In all the presented unification scenarios with light
precursors, various GUT scales are realized. This allows the model 
to be naturally embedded in either heterotic or Type ${\rm I}$ string
theories.

\end{abstract}
%\end{titlepage}

\section{Introduction}

One of the phenomenological motivations for SUSY GUTs is the successful
unification of the gauge couplings  near the scale $M_G\simeq
2\cdot 10^{16}$~GeV. Besides the nice unification picture SUSY
provides a natural understanding of the gauge hierarchy problem, but there
are puzzles which
need to be understood. Namely, one should find natural ways for
baryon number conservation and the resolution of the doublet-triplet
(DT) splitting
problem. The minimal $SU(5)$ and $SO(10)$ GUTs also suffer from the
problem of
wrong asymptotic mass relations: $\hat{M}_e^0=\hat{M}_d^0$ and
$\hat{M}_e^0=\hat{M}_d^0\propto \hat{M}_u^0$ for $SU(5)$ and $SO(10)$
respectively. Higher dimensional orbifold constructions 
suggest rather economical ways
for simultaneously resolving  these problems \cite{orbGUT1}-\cite{us}.
Therefore, they give  new insights for GUT model building\footnote{Proton 
stability and DT splitting in superstring derived models
were discussed in \cite{stringDT}.}.  
There are scenarios (not orbifold GUT constructions), which can give
unification near the TeV scale \cite{TEVuni}. However,
Let us note, that within orbifold GUTs only in some particular cases  
power law low scale unifications  is possible \cite{us}. While usually,
within 5D $S^{(1)}/Z_2\tm Z_2'$ GUTs, the unification is
logarithmic (like in 4D) and $M_G$ is still in the $10^{16}$~GeV
range \cite{orbGUTunif}, \cite{us}, with all GUT states lying  far above
the electro-weak
scale ($\sim 100$~GeV). This makes a test of most GUT models impossible in
present and future high energy colliders.

In a recent paper \cite{prec} by Dienes, Dudas and Gherghetta  
the possibility of an orbifold construction was suggested where the
compactification
scale $\mu_0=1/R$ ($R$ is the radius of the compact extra
dimension(s)) lies in the
TeV range, while the GUT scale is still near $10^{16}$~GeV. If GUT
symmetry
breaking occurs by boundary conditions upon compactification, then above
$\mu_0$ there appear
signatures of the GUT model: the KK modes of gauge bosons, which
correspond
to broken generators, will have masses $\stackrel{>}{_\sim }\mu_0$. In
\cite{prec}, these states were called GUT precursors, since their
appearance 
indeed would be a characteristic feature of the GUT model.

In this paper we present a detailed study of the possibility of light
precursors
coming from 5D GUTs compactified on an $S^{(1)}/Z_2\tm Z_2'$ orbifold. We
show that even if there is no power law relative running of gauge
couplings, there is a non universal logarithmic contribution which for
low
$\mu_0$ becomes significantly large and does not allow unification
with minimal field content. For unification, some extension
should be done. We present an extension of 5D SUSY SU(5), with additional
matter introduced on a brane, which can have unification in a range
$10^{10}-10^{16}$~GeV, depending on the selection of extended matter. 
We analyze also the 5D non SUSY $SU(5)$ model and it turns out that
also this one requires extensions, which seem to be rather complicated.
As an alternative GUT model, we study the extended GUT gauge group
$SU(6)$. As
it turns out, the non SUSY version of the 5D SU(6) GUT can naturally give
unification with light precursors. In all considered scenarios with
successful unification, the GUT scale can lie between scales
$10^{10}$~GEV and $10^{16}$~GeV, while the precursors have 
$\sim $TeV masses. These values of $M_G$ allow one to embed
the GUT scenario either in heterotic or in Type ${\rm I}$ string theory
(depending on which value for $M_G$ is realized).

\locsection{Renormalization for $S^{(1)}/Z_2\tm Z_2'$ orbifold 5D GUTs}

In this section we  present expressions, which will be useful for 
studying  gauge coupling unification for 5D GUTs.

As it was shown in \cite{orbGUT1}, a realistic 5D $SU(5)$ GUT can be built
if
compactification occurs on an $S^{(1)}/Z_2\tm Z_2'$ orbifold. For this
case, on one of the fixed points we have the 
$SU(3)_c\tm SU(2)_L\tm U(1)_Y\equiv G_{321}$ gauge group and minimal
field
content. If instead of $SU(5)$ some extended gauge group is considered,
compactification on an $S^{(1)}/Z_2\tm Z_2'$ orbifold still turns out to
be an economical possibility for realistic model building.

It is assumed that the fifth (space like) dimension $y$ parameterizes a
compact
$S^{(1)}$ circle with radius $R$. All states, introduced at 5D level,
should have definite $Z_2\tm Z_2'$ parities $(P, P')$, where
$Z_2:~ y\to -y$, $Z_2':~y'\to -y'$ ($y'=y+\pi R/2$). Therefore, there are
only the following options for parity prescription:
\beq
(P, P')=(+, +)~,~~(+, -)~,~~(-, +)~,~~(-, -)~,
\la{parpresc}
\eeq
and the corresponding KK states  have masses
\beq
2n \mu_0~,~~(2n+1)\mu_0~,~~(2n+1)\mu_0~,~~(2n+2)\mu_0~,
\la{KKmas}
\eeq
respectively, were  $\mu_0=1/R$ is the compactification scale and $n$ is
the
quantum number in the KK mode expansion. In a GUT scenario, 
KK states become relevant for gauge coupling running 
if $\mu_0$ lies below the GUT
scale $M_G$. The solution of one loop RGE has the form
\beq
\al_a^{-1}(M_G)=\al_a^{-1}(M_Z)-\fr{b_a}{2\pi }\ln \fr{M_G}{M_Z}+\De_a~,
\la{alGa}
\eeq
with
\beq
\De_a=\De_a^0+\De_a^{KK}~,
\la{Delta}
\eeq
where $b_a$ corresponds to the contribution of SM or MSSM states
(depending which case we consider) and
\beq
\De_a^0=-\fr{(b_a^{M_I})_{\al }}{2\pi }\ln \fr{M_G}{(M_I)_{\al }}~
\la{Delog}
\eeq
includes contributions from all additional brane and zero-mode 
states $\al $
with mass
$(M_I)_{\al }$. $\De_a^{KK}$ comes from the contributions of the KK states
(except zero-modes)
\beq
\De_a^{KK}=-\fr{\ga_a}{2\pi }S_1-\fr{\de_a}{2\pi }S_2~,
\la{DeKKZZ1}
\eeq
where $S_1$ and $S_2$ include contributions
from KK states with masses $(2n+2)\mu_0$ and $(2n+1)\mu_0$ resp.:
\beq
S_1=\sum_{n=0}^{N} \ln \fr{M_G}{(2n+2)\mu_0}~,~~~~
S_2=\sum_{n=0}^{N'} \ln \fr{M_G}{(2n+1)\mu_0}~.
\la{SKK}
\eeq
In (\ref{SKK}), $N$ and $N'$ are the maximal numbers of appropriate KK
states
which lie below $M_G$, e.g.
\beq
(2N+2)\mu_0\stackrel{<}{_\sim }M_G~,~~~~~
(2N'+1)\mu_0\stackrel{<}{_\sim }M_G~.
\la{maxNs}
\eeq
KK states with masses larger than $M_G$ are irrelevant.  
For a given
$M_G/\mu_0$ the $N$ and $N'$ can be calculated from (\ref{maxNs}).
Imposing the condition of gauge coupling unification
\beq
\al_1(M_G)=\al_2(M_G)=\al_3(M_G)\equiv \al_G~,
\la{coupuni}
\eeq
from (\ref{alGa}), eliminating $\al_G$ and $\ln M_G/M_Z$, we find for
the strong coupling at the $M_Z$ scale 
\beq
\al_3^{-1}=\fr{b_1-b_3}{b_1-b_2}\al_2^{-1}-
\fr{b_2-b_3}{b_1-b_2}\al_1^{-1}+
\fr{b_1-b_3}{b_1-b_2}\De_2-\fr{b_2-b_3}{b_1-b_2}\De_1-\De_3~,
\la{als}
\eeq
where $\al_a$ in (\ref{als}) stands for $\al_a(M_Z)$.
Also, from (\ref{alGa}) one can obtain 
\beq
\ln \fr{M_G}{M_Z}=\fr{2\pi }{b_1-b_2}(\al_1^{-1}-\al_2^{-1})+
\fr{2\pi }{b_1-b_2}(\De_1-\De_2)~,
\la{scale}
\eeq
and finally the value of the unified gauge coupling
\beq
\al_G^{-1}=\al_2^{-1}-\fr{b_2}{2\pi }\ln \fr{M_G}{M_Z}+\De_2~.
\la{alG}
\eeq

It is straightforward to present some expressions, which will also be useful
for further estimates.
{}For $N, N'\gg 1$ the $S_1$ and $S_2$ can be approximated by
using Stirling's formula
$$
S_1\simeq (N+1)\ln \fr{M_G}{\mu_0}-(N+\fr{3}{2})\ln (N+1)+
(1-\ln 2)(N+1)-\ln \sqrt{2\pi }-\fr{1}{12(N+1)}+\cdots ~,
$$
\beq
S_2\simeq (N'+1)\ln \fr{M_G}{\mu_0}-(2N'+\fr{3}{2})\ln (2N'+1)+
(N'+\fr{1}{2})\ln N'+(1+\ln 2)N'+1\cdots 
\la{apprS1S2}
\eeq
The combination, which will matter for the analysis below, is
\beq
S=S_2-S_1~,
\la{defS12}
\eeq
which according to (\ref{apprS1S2}), for $N=N'\gg 1$ reduces to the simple
form
\beq
S\simeq \ln \sqrt{\pi N}+\fr{7}{12N}+{\cal O}(\fr{1}{N^2})~.
\la{aprS12}
\eeq

\locsection{5D  $SU(5)$ GUT on $S^{(1)}/Z_2\times Z_2'$ orbifold}

We start our studies of  orbifold GUT models with a 5D 
$SU(5)$ theory. The fifth dimension is compact and  is considered to be
an
$S^{(1)}/Z_2\tm Z_2'$ orbifold. In terms of the $G_{321}$, the adjoint of 
$SU(5)$ reads 
\beq
{\bf{24}}=C(8,~1)_0+W(1,~3)_0+S(1,~1)_0+
X(3,~\bar 2)_{5}+Y(\bar 3,~2)_{-5}~,
\la{dec24ofSU5}
\eeq
where subscripts are the hypercharge 
$Y=\fr{1}{\sqrt{60}}\cdot {\rm Diag}(2, 2, 2, -3, -3)$ 
in the $1/\sq{60}$ units.
In the 5D gauge multiplet the 4D gauge field $A(24)$ is accompanied by
an adjoint
scalar $\Phi (24)$. In order to achieve $SU(5)\to G_{321}$ breaking, the
gauge fragments $A_X$, $A_Y$ [see decomposition in (\ref{dec24ofSU5})]
must have negative orbifold parity. This will insure the masses of their
KK modes to be proportional to $\mu_0$. Since the fragments $\Phi_X$,
$\Phi_Y$ should become the longitudinal modes of $A_X$, $A_Y$, the former
should carry opposite orbifold parity. Having only one $Z_2$, the
$\Phi_X$, $\Phi_Y$ would contain massless zero modes, which are
phenomenologically unacceptable. That's  why the projection on an 
$S^{(1)}/Z_2\tm Z_2'$ orbifold should be considered. Below we study SUSY
and non-SUSY cases separately.

\subsection{SUSY case}

The 5D $N=1$ SUSY gauge multiplet in 4D notation constitutes a $N=2$ SUSY
supermultiplet $V_{N=2}=(V, \Si )$, where $V$ is 4D $N=1$ gauge
supermultiplet and $\Si $ is a chiral superfield. 
Ascribing to the fragments [in decomposition (\ref{dec24ofSU5})]
of $V(24)$ and $\Si(24)$
the following $Z_2\tm Z_2'$ parities
$$
\l V_C,~V_W,~V_S\r \sim (+,~+)~,~~
\l V_X,~V_Y\r \sim (-,~+)~,
$$
\beq
\l \Si_C,~\Si_W,~\Si_S\r \sim (-,~-)~,~~
\l \Si_X,~\Si_Y\r \sim (+,~-)~,
\la{Z2Z2ofV24}
\eeq
at the $y=0$ fixed point we will have a 4D $N=1$ SUSY $G_{321}$ gauge
theory. We
introduce three families of quark-lepton superfields and two MSSM Higgs
doublets at the $y=0$ fixed point
[this $3$-brane is identified with our 4D
world]. We are looking for cases in
which $\mu_0 $ lies much below $M_G$; in order to have perturbativity
up to the GUT scale, we then always assume that matter and higgses are
introduced on the brane. This means that they do not have KK
excitations. Therefore,
the zero modes at the $y=0$ brane are just the MSSM content. 
Other states, including the $X$, $Y$ gauge bosons, are projected out.
Taking all this into account, and also (\ref{Z2Z2ofV24}),
(\ref{parpresc}), (\ref{KKmas}), for $b_a$ and $\ga_a$,
$\de_a$ [defined in (\ref{DeKKZZ1})] we have 
$(b_1, b_2, b_3)=(\fr{33}{5}, 1, -3)$,
\beq
\l \ga^{~}_1,~ 
\ga^{~}_2,~\ga_3\r =\l 0,~-4,~-6\r~,~~~~
\l \de_1,~\de_2,~\de_3\r =
\l -10,~-6,~-4 \r ~.
\la{BKKsu5} 
\eeq
{}From this and (\ref{als})-(\ref{alG}), taking also into account
(\ref{Delta})-(\ref{SKK}), we get for 'minimal' 5D SUSY $SU(5)$ the
following relations
\beq 
\al_3^{-1}=\fr{12}{7}\al_2^{-1}-\fr{5}{7}\al_1^{-1}-
\fr{3}{7\pi }S~,
\la{susysu5als} 
\eeq 
\beq 
\ln \fr{M_G}{M_Z}=\fr{5\pi }{14}(\al_1^{-1}-\al_2^{-1})+
\fr{5}{7}S~,
\la{susysu5scale} 
\eeq 
\beq 
\al_G^{-1}=\al_2^{-1}-\fr{1}{2\pi }\ln \fr{M_G}{M_Z}+
\fr{2}{\pi }S_1+\fr{3}{\pi }S_2~.
\la{susysu5alG}  
\eeq
Here, (\ref{susysu5als}) and (\ref{susysu5scale}) show that $\al_3$ and
$M_G$
are determined by the function $S$, which is defined in (\ref{defS12}).
{}For simplicity let's take $N=N'$, which for large values allows to
use  approximation (\ref{apprS1S2}). With this, from (\ref{susysu5als}) we
see that already  $N=N'=10^6$ gives an unacceptably large 
$\al_3(M_Z)$($\simeq 0.13$), while (\ref{susysu5scale}) gives an increased
value of the GUT scale($\simeq 4\cdot 10^{18}$~GeV).
The situation becomes even worse for larger $N$, $N'$, because the 
function $S$ grows logarithmically. Therefore, we conclude
that, 
in this minimal setting, the compactification scale 
$\mu_0\simeq M_G/(2N)$
can not be lowered below $4\cdot 10^{12}$~GeV. 
The first two terms in (\ref{susysu5als}) would give a nice value for 
$\al_3$($\simeq 0.116$), which coincides with the one loop 4D $SU(5)$
prediction. To maintain this, one should take $S\simeq 0$, which means
$\mu_0\simeq M_G$. This excludes light
precursors in the framework of minimal 5D SUSY $SU(5)$.

\hs{-0.6cm}{\bf Light precursors from extended 5D SUSY SU(5)}

\vs{0.1cm}

\hs{-0.6cm}With specific extension of minimal 5D SUSY $SU(5)$, it is 
possible to get
successful unification with low $\mu_0$. Let's at the $y=0$ brane
introduce the following vector like states
\beq
N_E \tm (E^c+\ov{E}^c)~,~~~~N_L\tm (L+\ov{L})~,
\la{extst}
\eeq
where under $G_{321}$, $E^c$ and $L$  have precisely the same
transformation
properties as a right handed charged lepton and a left handed lepton
doublet
respectively. $\ov{E}^c$, $\ov{L}$ have conjugate quantum numbers and
$N_E$, $N_L$ denote the numbers of corresponding vector like
pairs. Assuming
that they have 4D masses $M_E$, $M_L$, above these scales these states
will contribute to the $b$-factors as
\beq
\l b_1, b_2, b_3\r^E=\l \fr{6}{5}, 0, 0 \r \cdot N_E~,~~~~
\l b_1, b_2, b_3\r^L=\l \fr{3}{5}, 1, 0 \r \cdot N_L~.
\la{extb}
\eeq
In this case, the RGEs are modified and have the form
\beq 
\al_3^{-1}=\fr{12}{7}\al_2^{-1}-\fr{5}{7}\al_1^{-1}
+\fr{3N_E}{7\pi }\ln \fr{M_G}{M_E}-\fr{9N_L}{14\pi }\ln \fr{M_G}{M_L}
-\fr{3}{7\pi }S~,
\la{extsusysu5als} 
\eeq 
\beq 
\ln \fr{M_G}{M_Z}=\fr{5\pi }{14}(\al_1^{-1}-\al_2^{-1})
-\fr{3N_E}{14\pi }\ln \fr{M_G}{M_E}+\fr{N_L}{14\pi }\ln \fr{M_G}{M_L}
+\fr{5}{7}S~,
\la{extsusysu5scale} 
\eeq 
\beq 
\al_G^{-1}=\al_2^{-1}-\fr{1}{2\pi }\ln \fr{M_G}{M_Z}
-\fr{N_L}{2\pi }\ln \fr{M_G}{M_L} 
+\fr{2}{\pi }S_1+\fr{3}{\pi }S_2~.
\la{extsusysu5alG}  
\eeq
{}From (\ref{extsusysu5als}), (\ref{extsusysu5scale}) we see that by
suitable selections of $N_E$, $N_L$ and of the mass scales, it is possible
to
cancel the large logarithmic contribution coming from $S$ and obtain
a reasonable
value for $\al_3(M_Z)$. At the same time, it is possible to get various
values of $M_G$. The different cases of unification, estimated through
(\ref{extsusysu5als}), (\ref{extsusysu5scale}), are presented in Table
\ref{t:extsusysu5}. As we see, for relatively large $N_E$ and $N_L$, it is
also possible to have a lower unification scale. For the considered cases
the $\mu_0$ lies in the TeV range.

\begin{table} \caption{Extended 5D SUSY $SU(5)$ with light precursors. For
all cases $\al_3(M_Z)=0.119$.}
 
\label{t:extsusysu5} $$\begin{array}{|c|c|c|c|c|c|c|c|}
 
\hline 
{\rm Case }&N,~ N'&N_E &N_L & M_G/M_E & M_G/M_L&
M_G/{\rm GeV} & \mu_0   \\
 
\hline \hline
 
{\rm (I)} &10^{13} &2 &1 &3\cdot 10^{13} &3\cdot 10^{13} &
2\cdot 10^{16}  &1 {\rm TeV} \\ 

\hline

{\rm (II)} &10^{10} &4 &2 &6.4\cdot 10^9 &2.1\cdot 10^{11} &
1.8\cdot 10^{13} &912~ {\rm GeV} \\ 

\hline  
{\rm (III)} &10^{10} &4 &3 &6.4\cdot 10^9 &3.6\cdot 10^7 &
1.8\cdot 10^{13} &912~ {\rm GeV} \\

\hline  

\end{array}$$
 
\end{table}

\subsection{Non-SUSY case}

In this subsection, we study the possibility of light precursors for 
the non-SUSY case. For the non supersymmetric model, the higher
dimensional
extension is more straightforward than for the SUSY one. As was pointed
out at
the beginning of sect. 3, in a 5D extension the 4-dimensional gauge field
should be accompanied by a real scalar. As far as the matter and Higgs
fields are concerned, we introduce them on the brane. The 
$S^{(1)}/Z_2\tm Z_2'$ orbifold parities for bulk states still are chosen
in such a way as to break $SU(5)$ down to the $G_{321}$. It is easy to
check  that also for minimal 5D non-SUSY $SU(5)$, the 
light precursors are incompatible with unification due to large
logarithmic corrections [caused by  $S$ of (\ref{defS12})] coming from
bulk states.

\begin{table} \caption{Extended 5D non-SUSY $SU(5)$ with light
precursors. For
all cases $\al_3(M_Z)=0.119$.}

\label{t:extsu5} $$\begin{array}{|c|c|c|c|c|c|c|c|}
 
\hline 
{\rm Case }&N,~ N'&N_E &N_L & M_G/M_E & M_G/M_L&
M_G/{\rm GeV} & \mu_0   \\
 
\hline \hline
 
{\rm (I)} &5\cdot 10^{11} &8 &4 &2.5\cdot 10^{13} &1.2\cdot 10^{10} &
3.7\cdot 10^{16}  &37 {\rm TeV} \\ 

\hline

{\rm (II)} &5\cdot 10^{9} &11 &5 &7.4\cdot 10^{10} &3.3\cdot 10^{10} &
3.4\cdot 10^{13} &3.4~ {\rm TeV} \\ 

\hline  
{\rm (III)} &10^{8} &14 &7 &9.5\cdot 10^8 &4.6\cdot 10^8 &
2.3\cdot 10^{11} &1.1~ {\rm TeV} \\

\hline  

\end{array}$$
 
\end{table}

\vs{0.1cm}

\hs{-0.6cm}{\bf Light precursors from extended 5D non-SUSY SU(5)}

\vs{0.1cm}

\hs{-0.6cm}Also here we extend the model by introducing (at the $y=0$
brane) vector like states with the transformation properties of
(\ref{extst}),
but now instead of superfields, under $\ov{E}^c$, $E^c$, $\ov{L}$, $L$ we
assume fermionic states. Their contribution to the $b$-factors are
%\beq
%{\rm Vector~like~ fermions}~~N_E \tm (E^c+\ov{E}^c)~,~~~N_L\tm
%(L+\ov{L})~,
%\la{extstsu5}
%\eeq
\beq
\l b_1, b_2, b_3\r^E=\l \fr{4}{5}, 0, 0 \r \cdot N_E~,~~~~
\l b_1, b_2, b_3\r^L=\l \fr{2}{5}, \fr{2}{3}, 0 \r \cdot N_L~.
\la{extbsu5}
\eeq
With this setting the one loop solutions of RGEs have the forms
\beq 
\al_3^{-1}=\fr{333}{218}\al_2^{-1}-\fr{115}{218}\al_1^{-1}
+\fr{23N_E}{109\pi }\ln \fr{M_G}{M_E}-
\fr{44N_L}{109\pi }\ln \fr{M_G}{M_L}
-\fr{21}{218\pi }S~,
\la{extsu5als} 
\eeq 
\beq 
\ln \fr{M_G}{M_Z}=\fr{30\pi }{109}(\al_1^{-1}-\al_2^{-1})
-\fr{12N_E}{109\pi }\ln \fr{M_G}{M_E}+\fr{4N_L}{109\pi }\ln \fr{M_G}{M_L}
+\fr{105}{109}S~,
\la{extsu5scale} 
\eeq 
\beq 
\al_G^{-1}=\al_2^{-1}+\fr{19}{12\pi }\ln \fr{M_G}{M_Z}
-\fr{N_L}{3\pi }\ln \fr{M_G}{M_L} 
+\fr{7}{2\pi }S_1+\fr{21}{4\pi }S_2~.
\la{extsu5alG}  
\eeq
{}From them it follows that one can get successful unification
with
light precursors. Cases with different mass scales and number of
vector like states are presented in Table \ref{t:extsu5}. As we see, the
$N_E$, $N_L$ should be large. This indicates that an extension is
complicated.

\locsection{5D  SU(6) GUT on $S^{(1)}/Z_2\tm Z_2'$ orbifold} 

In the previous section we have seen that in order to have light
precursors within 5D $SU(5)$ GUTs, one has to extend the matter
sector. We will now discuss the extension of the gauge group. The smallest
unitary group (in rank), which includes $SU(5)$ is an $SU(6)$ 
and we will consider here its 5D gauge version. The symmetry breaking of
$SU(6)$ should occur in two stages. Through compactification, by proper 
selection of the $S^{(1)}/Z_2\tm Z_2'$ orbifold parities, the $SU(6)$ can
be 
reduced to one of its subgroups ${\cal H}$. For the latter, there are
three possibilities:
${\cal H}={\cal H}_{331}=SU(3)_c\tm SU(3)_L\tm U(1)$,
${\cal H}={\cal H}_{421}=SU(4)_c\tm SU(2)_L\tm U(1)$ and
${\cal H}={\cal H}_{51}=SU(5)\tm U(1)$.
The ${\cal H}$ gauge symmetry is realized at the $y=0$ fixed point and its
further breaking
to the $G_{321}$ must occur spontaneously, by the VEV of an appropriate
scalar
field. As it turns out, the cases of 
${\cal H}={\cal H}_{421}~ {\rm or}~ {\cal H}_{51}$ do
not allow for light precursors, while the non-SUSY case of 
${\cal H}={\cal H}_{331}$ gives an interesting possibility. The 5D SUSY
SU(6) with minimal field content does not lead to unification with light
precursors. Of course, some extensions of  SUSY $SU(6)$ versions in the
matter sector can give the desirable result, but since we are looking for
an $SU(6)$ model with minimal matter/Higgs content, we will not pursue 
this possibility and concentrate on the non-SUSY version.

\vs{0.1cm} 
\hs{-0.6cm}{\bf Light precursors from 5D non-SUSY  SU(6)} 
\vs{0.1cm}

\hs{-0.6cm}The adjoint representation ${\bf 35}$ of the $SU(6)$, in terms
of 
${\cal H}_{331}=SU(3)_c\tm SU(3)_L\tm U(1)$ decomposes as
\beq
{\bf 35}=C(8, 1)_0+L(1, 8)_0+S(1, 1)_0+C\ov{L}(3, \bar 3)_2+
\ov{C}L(\bar 3, 3)_{-2}~,
\la{dec35}
\eeq
where the subscripts denote $U(1)$ hypercharges
\beq
Y_{U(1)}=\fr{1}{\sq{12}}\l 1,~1,~1,~-1,~-1,~-1 \r~,
\la{Ysu6}
\eeq
in $1/\sq{12}$ units (the $SU(6)$ normalization).
Choosing the $Z_2\tm Z_2'$ parities of the fragments of $A_{\mu}(35)$ and
$\Phi (35)$ [together they form a 5D gauge field ${\bf A}=(A, \Phi
)$] 
as
$$
\l A_C, A_L, A_S\r \sim (+, +)~,~~~ 
\l A_{C\bar L}, A_{\bar C L}\r\sim (-, +)~, 
$$ 
\beq 
\l \Phi_C, \Phi_L, \Phi_S\r \sim (-, -)~,~~~ 
\l \Phi_{C\bar L}, \Phi_{\bar C L}\r\sim (+, -)~, 
\la{SU6gaugepar} 
\eeq 
at the $y=0$ fixed point we will have ${\cal H}_{331}$ gauge symmetry. For
its
breaking down to $G_{321}$,  we introduce a Higgs field $H$ at the $y=0$
brane
transforming under ${\cal H}_{331}$ as $(1, 3)_{-1}$. Its third
component's non zero VEV $\lan H\ran \equiv v$ induces the breaking
$SU(3)_{L}\tm U(1)\stackrel{v}{\longrightarrow}SU(2)_L\tm U(1)_Y $, where
\beq
Y=-\fr{1}{\sqrt{5}}Y_{U(1)'}+\fr{2}{\sqrt{5}}Y_{U(1)}~,
\la{supcharges}
\eeq
and $Y_{U(1)'}$ is the $SU(3)_L$ generator
\beq
Y_{U(1)'}=\fr{1}{\sqrt{12}}{\rm Diag }\l 1, 1, -2\r~.
\la{Ycharges}
\eeq
At the $y=0$ fixed point we also introduce three families of matter
\beq
u^c(\ov{3},~1)_2~,~~Q(3,~3)_0~,~~E^c(1,~\ov{3})_{-2}~,~~
d^c_{1, 2}(\ov{3},~1)_{-1}~,~~{\cal L}_{1, 2}(1,~\ov{3})_1~,
\la{SU6matter}
\eeq
where we marked the transformation properties under ${\cal H}_{331}$.
Under $G_{321}$
\beq 
Q=(q,~\ov{D}^c)~,~~~E^c=(e^c,~\ov{L})~,~~~{\cal L}=(l,~\xi ) ~,
\la{decQEL} 
\eeq 
where $\xi $ is a SM singlet. It is easy to verify that
(\ref{SU6matter}) effectively constitute anomaly free $15+\bar 6_{1, 2}$
chiral multiplets of the $SU(6)$ gauge group.
Together with these, at the $y=0$ brane, we introduce an $h(1, 3)_{-1}$
Higgs,
containing the SM Higgs doublet. With the brane couplings
$Qd^c_{1, 2}H^{+}$, $E^c{\cal L}_{1, 2}H^{+}$, after substituting the
$H$'s VEV $v$, the extra $\ov{D}^c$ and $\ov{L}$ states form massive
states with one
superposition of $d^c_{1, 2}$ and $l_{1, 2}$ respectively and therefore
decouple. So, below the
scale $v$, we have the SM with its minimal content. Taking 
all this into
account, above the scale $v$, the $b$-factors are
\beq
\l b_1, b_{3L}, b_{3}\r =\l \fr{37}{6}, -1, -5\r~,
\la{bsu6}
\eeq
while taking into account (\ref{SU6gaugepar}) the 
$\ga $ and $\de $-factors
read 
\beq
\l \ga_1, \ga_{3L}, \ga_3\r =
\l 0, -\fr{21}{2}, -\fr{21}{2} \r~,~~~
\l \de_1, \de_{3L}, \de_3\r =
\l -21, -\fr{21}{2}, -\fr{21}{2}\r~.
\la{gadesu6}
\eeq
{}From (\ref{supcharges}), it follows that
\beq
\ga_Y=-\fr{21}{10}~,~~~\de_Y=
-\fr{189}{10}~,~~~b_Y=\fr{71}{15}~.
\la{factY}
\eeq
Using (\ref{bsu6})-(\ref{factY}) we derive
\beq 
\al_3^{-1}=\fr{333}{218}\al_2^{-1}-\fr{115}{218}\al_1^{-1}
-\fr{319}{654\pi }\ln \fr{M_G}{v}
-\fr{483}{218\pi }S~,
\la{su6als} 
\eeq 
\beq 
\ln \fr{v}{M_Z}=\fr{30\pi }{109}(\al_1^{-1}-\al_2^{-1})
-\fr{215}{218\pi }\ln \fr{M_G}{v}
-\fr{63}{109}S~,
\la{su6scale} 
\eeq 
%\beq 
%\al_G^{-1}=\al_2^{-1}-\fr{1}{2\pi }\ln \fr{M_G}{M_Z}
%-\fr{N_L}{2\pi }\ln \fr{M_G}{M_L} 
%+\fr{2}{\pi }S_1+\fr{3}{\pi }S_2~.
%\la{su6alG}  
%\eeq
{}From (\ref{su6als}) we already see, that with help of the last two
terms,
the wrong prediction of minimal non-SUSY $SU(5)$ 
[$(\al_3^{-1})^{\rm min}_{SU(5)}=
\fr{333}{218}\al_2^{-1}-\fr{115}{218}\al_1^{-1} \simeq 1/0.07$] can be
improved. Namely,
for $N=N'=3\cdot 10^6$, $M_G/v=1.27$ we obtain successful unification
with $\al_3(M_Z)=0.119$, $v\simeq 7\cdot 10^{10}$~GeV. Therefore,
unification occurs at $M_G\simeq 8.8\cdot 10^{10}$~GeV and the
compactification scale is 
$\mu_0=14.7$~TeV, i.e. we have light precursors.

\section{Discussions and conclusions }

In this paper we have presented a detailed study of the possibility of
light
precursors within concrete 5D GUTs. We have shown that compactification on
a $S^{(1)}/Z_2\tm Z_2'$ orbifold introduces corrections, which
significantly affect gauge coupling unification. In order to compensate
these corrections, specific extensions are needed. Achieving successful
unification with light precursors, the prediction for $\al_3(M_Z)$ will
not be affected by possible brane localized gauge kinetic terms, since the
bulk is sufficiently large. Although the relative running of gauge
couplings is logarithmic, the coupling itself has power law
running. According
to (\ref{apprS1S2}), in the ultraviolet limit ($N\simeq N'\to \infty $) 
we have $S_1\simeq S_2\to (1-\ln 2)N$ and the dominant contribution to
the
renormalization (\ref{alGa}) comes from KK states:
$\al_a^{-1}\to -\fr{\ga_a+\de_a}{2\pi }(1-\ln 2)N$. For a given GUT,
$\ga_a+\de_a={\rm const.}\equiv \tl{b}$ and the effective coupling is
\cite{prec} $\al_a^{\rm eff}=(1-\ln 2)N\al_a=-\fr{2\pi }{\tl{b}}$.
Indeed $\al_5^{-1}\sim (\al^{\rm eff})^{-1}\La $, where $\La $ is the
ultraviolet cut off and $\al_5$ is the 5D gauge coupling.
The $\al^{\rm eff}$ remains perturbative if $\tl{b}$ is large and
negative: 
$-\fr{2\pi}{\tl{b}}\ll 4\pi$. It is easy to check that, in all models
considered above, this condition is well satisfied. Therefore, at the
ultraviolet limit $\al^{\rm eff}$ approaches a perturbativly fixed value.

As far as the fundamental scale $M_*$ (5D Planck scale) is concerned, with
the simplest setting the four and five dimensional Planck scales are
related as
\cite{ADD} $M_{\rm Pl}^2\sim M_*^3R$ and for $1/R\sim $~TeV we have 
$M_*\sim 10^{14}$~GeV. To have a self-consistent picture of unification,
we
need the unification scale $M_G$ to lie below the $M_*$. Attempting to
embed the presented scenarios into string theory, one also should make the
values of scales self-consistent\footnote{For detailed discussions see
\cite{prec}.}. In perturbative heterotic string theory 
$M_{\rm string}\sim M_*$ and there is a similar relation 
($M_{\rm Pl}^2\sim M_*^3R$).
Therefore, $M_G$ still should lie below the $10^{14}$~GeV.
However, within the context of Type ${\rm I}$ (non perturbative) strings,
$M_*$ can be close to $10^{16}$~GeV, which allows to have unification in
this region. Within the scenarios considered in this paper, it is
possible to have $M_G$ in the range of $(10^{11}-10^{16})$~GeV. This
allows the models to be embedded either in heterotic or in 
Type ${\rm I}$ string theory.

\vspace{0.3cm} 
\hs{-0.6cm}{\bf Acknowledgements} 

\hs{-0.6cm}Research of F.P.C. is supported by 
Funda\c c\~ ao para a Ci\^ encia e a
Tecnologia
(grant SFRH/BD/4973/2001).

\vspace{0.2cm} 

{\it Note added}: After the submission of this paper into the arXive, we
were informed by authors of \cite{prec}, that the presence of additional
logarithmic contributions into the relative gauge coupling runnings was 
reported  in the Pascos'03 meeting by E. Dudas \cite{pascos}.

%\end{document}

\bibliographystyle{unsrt}

\end{document}